\documentclass[10pt,twocolumn]{article} 
\usepackage{amssymb}
\usepackage{ol2}
\usepackage{amsmath}

\begin{document}

\twocolumn[ 
OPTICS LETTERS, {\it submitted} Sept. $16^{th}$, 2010, {\it revised} Dec. $13^{th}$, 2010,  manuscript ID: 135301

\title{All-fiber controller of radial polarization using a periodic stress}
\author{Djamel Kalaidji $^{1,2}$, Michel Spajer $^{1,*}$, Thierry Grosjean $^{1}$}

\address{$^1$ Institut FEMTO-ST, Universit\' e de Franche-Comt\' e, \\ UMR 6174 CNRS, 32 avenue de l'Observatoire, 25044 Besan\c con cedex, France \\
$^2$ Universit\' e Abou Bekr Belkaid, Facult\' e des Sciences, \\ D\' epartement de Physique, BP 119, 13000 Tlemcen, Alg\' erie \\
$^*$ Corresponding author: michel.spajer@univ-fcomte.fr
}

\begin{abstract}
Our aim is to transpose the polarization control by mechanical stress usually applied to single mode fibers, to the ($TM_{01}, TE_{01}, HE_{21}^{ev}, HE_{21}^{od}$) annular mode family. Nevertheless, the quasi degeneracy of these four modes makes the situation more complex than with the fundamental mode $HE_{11}$. We propose a simple device based on periodic perturbation and mode coupling to produce the radially polarized $TM_{01}$ mode or at least one single of the four modes at the extremity of an arbitrarily long fiber, the conversion to $TM_{01}$ mode being achievable by classical crystalline plates.

\end{abstract}


] 

\maketitle
\newpage
\noindent  Two properties of annular or doughnut beams \cite{zhan09} have been exploited for various applications, such as vortex generation or specific polarization for microscopy.  A method was proposed \cite{grosjean05} to obtain a perfectly radial polarization by selecting the $TM_{01}$ mode from a short segment of a two-mode fiber. We attempt to generalize the process to obtain the desired polarization after a propagation along several meters. This paper deepens the understanding of mode selectivity and proposes an experimental verification based on two principles: a) the theory of mode coupling related to periodic perturbations \cite{snyder83} and the analogy with other mode couplers \cite{youngquist84}, the adapted coupling transforming the injected mode $LP_{11}$ into one of its components, $TM_{01}$ or $HE_{21}$, b) a rigorous treatment of the propagation modes in an anisotropic fiber \cite{alekseev02} whose eigenmodes are a combination of the four modes $TM_{01}$, $TE_{01}$, $HE_{21}^{ev}$ and $HE_{21}^{od}$ of the isotropic fiber. The mode nomenclature is explained by equation (2), their respective field strength will be noted $(HE_{21}^{od})$, etc ... and their normalized power $|(HE_{21}^{od})|^2$, etc ...

\begin{figure}[!htbp]
        \includegraphics[width=8.5cm]{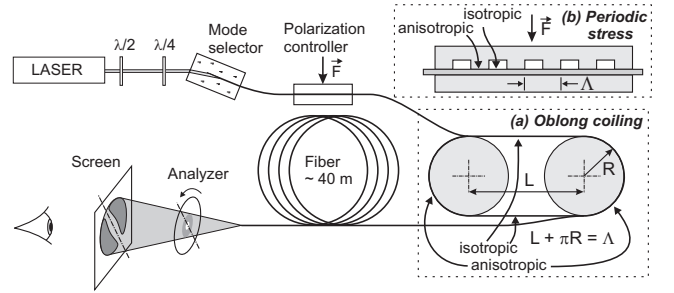}
        \caption{Injection and coupling device: the oblong coiling (a) that induces the periodic stress replaces the usual series of ridges (b).}
       \label{Setup}
\end{figure}

Among the all-fiber solutions, an efficient beam converter was based on a short length of stressed two-mode fiber \cite{mcgloin98} but its aim was a doughnut beam with a linear azimuthal phase regardless of the polarization. Several theoretical publications demonstrated the conditions of stability of vortices in anisotropic, twisted or coiled fibers \cite{alexeyev08}\cite{alexeyev07b}. Besides, a periodic diametral compression is known to couple $LP_{11}$ and $LP_{01}$ modes if the period equals the beat length $\Lambda$ (typ. 0.3 mm) of the two modes \cite{youngquist84}. In the same way, $TM_{01}$ and $HE_{21}^{ev}$ modes can be coupled by a stress of long spatial period (typ. 170 mm), as the two modes have almost the same propagation constant. Injection of $LP_{11x}^{ev}$ mode \cite{grosjean05} can be considered as the simultaneous injection of $TM_{01}$ and $HE_{21}^{ev}$ modes. The device that will couple the whole energy into one of these modes can keep a reasonable size ($\approx 150 mm$, 4 times longer than the $LP_{01}/LP_{11}$ converters \cite{youngquist84}) since coiling the fiber on a non-circular cylinder produces the desired periodic perturbation. We chose two cylinders of radius $R$ separated by a distance $L$ between which the coil is stretched. Fig.\ref{Setup} shows the analogy between a periodic stress and this oblong coiling: one period of the perturbation is made of a straight segment and half a circle. Its total length must equal the beat length to induce an efficient coupling between the two modes. One of the cylinders is mounted on a translation stage to control the stretching of the fiber.
Straight and curved segments have different eigenmodes: the 4-coefficient decomposition of the transverse field in the curved segment is derived from the other by the $4 \times 4$ matrix $M$ written in (1):
\begin{equation}
{\fontsize{8.5}{10}
\begin{vmatrix}
(\Psi _1)\\(\Psi _2)\\(\Psi _3)\\(\Psi _4)
\end{vmatrix}=\frac{1}{\sqrt{2}} \begin{array}{|l|}
\cos \Theta _1 \hspace{1 mm} - \sin \Theta _1 \hspace{4 mm} 0 \hspace{9 mm} 0\\
\sin \Theta _1 \hspace{3 mm} \cos \Theta _1 \hspace{5 mm} 0 \hspace{9 mm} 0\\
\hspace{3 mm} 0 \hspace{9 mm} 0 \hspace{5 mm} \sin \Theta _3 \hspace{3 mm} \cos \Theta _3\\
\hspace{3 mm} 0 \hspace{9 mm} 0 \hspace{5 mm} \cos \Theta _3 \hspace{1 mm} - \sin \Theta _3
\end{array}
\begin{vmatrix}
(HE_{21}^{od}) \\ (TE_{01}) \\ (HE_{21}^{ev}) \\ (TM_{01})
\end{vmatrix}
}
\end{equation}
This equation is deduced from the description by Alekseev {\it et al.} \cite{alekseev02}(Table 1) of the eigenmodes $\Psi _i$ in an anisotropic fiber. Their transverse field component can be written:
\begin{equation}
{\fontsize{8.5}{10}
\begin{array}{lll}
\theta _i = \frac{\pi}{4} & \pi /4 \geq \theta _i \geq 0 & \theta _i = 0 \\
HE_{21}^{od} & \Psi_1 = {\bf e}_x \cos \theta _1 \sin \phi + {\bf e}_y \sin \theta _1 \cos \phi  & LP_{11x}^{od} \\
TE_{01} & \Psi_2 = - {\bf e}_x \sin \theta _1 \sin \phi + {\bf e}_y \cos \theta _1 \cos \phi  & LP_{11y}^{ev} \\
HE_{21}^{ev} & \Psi_3 = {\bf e}_x \cos \theta _3 \cos \phi - {\bf e}_y \sin \theta _3 \sin \phi  & LP_{11x}^{ev} \\
TM_{01} & \Psi_4 = {\bf e}_x \sin \theta _3 \cos \phi + {\bf e}_y \cos \theta _3 \sin \phi  & LP_{11y}^{od}
\end{array}
}
\end{equation}
where ${\bf e}_x(r)$ and ${\bf e}_y(r)$ are respectively x- and y- polarized field distributions with the same doughnut r-dependence, $\phi$ is the azimuthal angle and $\theta _i = \pi /4 - \Theta _i$ is derived from the equations:
\begin{equation}
A = \frac{2\Delta}{r_0^2}\frac{u w^2}{V^2} \frac{J_1(u)}{J_0(u)}, \hspace{2 mm} B = - \frac{4\Delta}{r_0^2}\frac{u w^2}{V^2} \frac{J_1(u)}{J_2(u)}
\end{equation}
{\fontsize{9}{10}
\begin{equation}
E_1 = n_0 k^2 \delta n, \hspace{1 mm} \tan (2\theta _1) = \frac{A}{2 E_1}, \hspace{1 mm} \tan (2\theta _3) = \frac{A+ 2 B}{2 E_1}
\end{equation}
We used a He-Ne laser @ 633 nm and a standard step index fiber with the following parameters: core radius $r_0 = 2.02 \mu m$, core index $n_0=1.47$, normalized index step $\Delta = 0.0053$, normalized frequency $V=3.037$, scale parameters $u=2.733$ and $w=1.325$. As the geometrical birefringence can be neglected \cite{garth87} for moderate curvatures, the anisotropy $\delta n=n_x-n_y$ of a curved fiber under a tension $T$ is \cite{ulrich80}\cite{ulrich80b}, with the usual photoelastic coefficients:
\begin{eqnarray}
\delta n = 0.135  (b^2 / R^2) + 0.492 (b/R) \epsilon _z
\end{eqnarray}
where $b=62.5 \mu m$ is the fiber radius, $R$ its radius of curvature, $\epsilon _z = T / E \pi b^2$ its longitudinal deformation and $E$ its Young modulus ($E=7.33 \hspace{1 mm} 10^{10} N/m^2$).

The mode propagation is described by a diagonal matrix $M_L$ ($d = L$, $E_1 = 0$) for the straight segment, $M_R$ ($d = \pi R$, $E_1 \ge 0$) for the curved one:
\begin{equation}
\begin{matrix}
M_R \\
M_L
\end{matrix}=
\begin{vmatrix}
e^{-j \beta _1 d} & 0 & 0 & 0\\
0 & e^{-j \beta _2 d} & 0 & 0\\
0 & 0 & e^{-j \beta _3 d} & 0\\
0 & 0 & 0 & e^{-j \beta _4 d}
\end{vmatrix}
\end{equation}
The propagation constants $\beta _i=\beta + \Delta \beta_i$ are deduced from the constant $\beta$ of the scalar approximation as follows \cite{alekseev02}(Table 1):
\begin{equation}
\begin{array}{ll}
\Delta \beta_1 = & \frac{1}{4 \beta_{1}}[A - (A^2 + 4 E_{1}^2)^{1/2}] \\
\Delta \beta_2 = & \frac{1}{4 \beta_{2}}[A + (A^2 + 4 E_{1}^2)^{1/2}] \\
\Delta \beta_3 = & \frac{1}{4 \beta_{3}}[A + 2 B + ((A-2B)^2 + 4 E_{1}^2)^{1/2}] \\
\Delta \beta_4 = & \frac{1}{4 \beta_{4}}[A + 2 B - ((A-2B)^2 + 4 E_{1}^2)^{1/2}]
\end{array}
\end{equation}
This equation contains the usual polarization correction for an ideal straight fiber when $E_{1}=0$ \cite{snyder83}(Table 14-6: $\Delta \beta_1 = \Delta \beta_3 = \Delta \beta_{HE}, \Delta \beta_2 = \Delta \beta_{TE} = 0, \Delta \beta_4 = \Delta \beta_{TM}$) and its modification by the curvature anisotropy ($E_{1} \neq 0$).
Therefore the transfer of the mode distribution by one period and $n$ periods of the device, if we suppose arbitrarily that the curved segment comes first, are respectively characterized by the matrix $T$ and $T^n$ :
\begin{equation}
T= M_L M^{-1} M_R M, \hspace{3 mm} T^n= (M_L M^{-1} M_R M)^n
\end{equation}
This formalism implicitly neglects two phenomena that induce a circular birefringence: Berry's phase \cite{alexeyev07b}(eq.8) and twist \cite{alexeyev08}(eq.19). Both have no effect on $TE$ and $TM$ modes but give rise to a rotation of $HE$ modes, or in other words a difference in the propagation constants of the two circularly polarized modes that compose them, respectively $\Delta \beta _B$ and $\Delta \beta _{tw}$. They can eventually compensate one another, or $\Delta \beta _B$ can be canceled by maintaining the curved segment in a horizontal plane, the coiling being helped by a slight slope of the straight segments. We skipped these delicate adjustments because $\Delta \beta _B$ is only $1.5 \%$ of $\Delta \beta _3 - \Delta \beta _{HE}$, i.e. the effect of linear birefringence. Therefore only the diagonal quadrants of matrix $T$ and $T^n$ have non-zero terms, hence the coupling processes in each sub-system, ($TM_{01}$, $HE_{21}^{ev}$) and ($TE_{01}$, $HE_{21}^{od}$), do not interfere, so $TM_{01}$/$HE_{21}^{ev}$ coupling is fully defined by the fourth quadrant of $T^n$, i.e. a $2 \times 2$ matrix.
\begin{figure}[!htbp]
        \includegraphics[width=8.5cm]{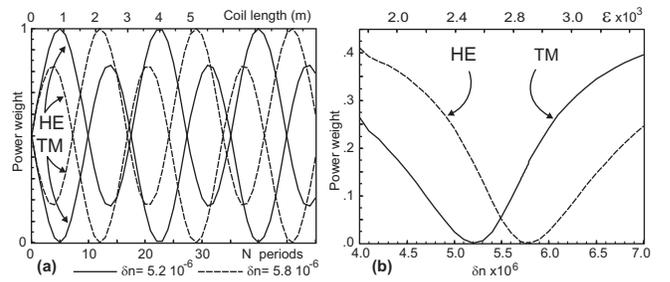}
        \caption{Intermodal coupling: (a) normalized power of $HE_{21}^{ev}$ and $TM_{01}$ vs. coiling length for R = 20 mm, $\Lambda$ = 177.6 mm and 2 values of the birefringence $\delta n$, (b) minimum of $|(HE_{21}^{ev})|^2$ and $|(TM_{01})|^2$ vs. $\delta n$}
       \label{Coupling}
\end{figure}
Fig.\ref{Coupling} models the evolution of $|(TM_{01})|^2$ and $|(HE_{21}^{ev})|^2$ for a particular geometry of the coiling. At the fiber input the energy of $LP_{11}$ mode is equally distributed on both annular modes. The results of Fig.\ref{Coupling}(a) are obtained with a tension that cancels $TM_{01}$ mode at the $5^{th}$ period and an other one that cancels $HE_{21}^{ev}$ mode at the $12^{th}$ period. The variation of the minimum of $|(HE_{21}^{ev})|^2$ and $|(TM_{01})|^2$ with birefringence $\delta n$ and deformation $\epsilon _z$ is detailed by Fig.\ref{Coupling}(b). The main result is the small number of periods that are necessary to ensure the coupling: a five period coil has been chosen for our experimental measurements.
\begin{figure}[!htbp]
        \includegraphics[width=8.5cm]{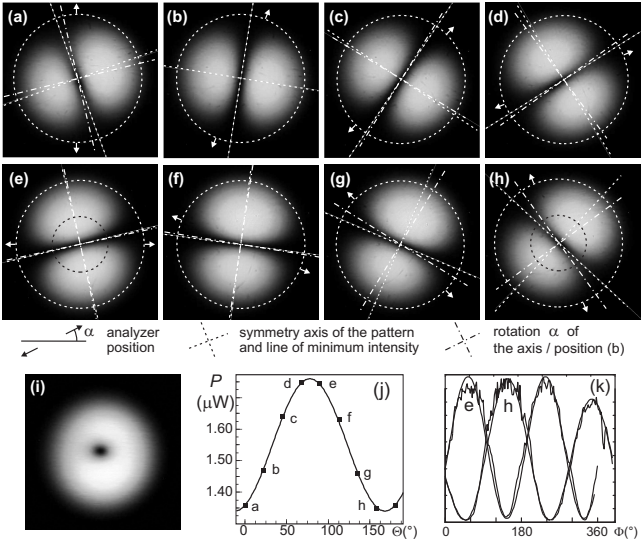}
        \caption{Mode pattern vs. analyzer rotation: (a-h) 8 analyzer positions, (i) without analyzer, (j) power transmitted by the analyzer, (k) azimuthal variation of intensity along the inner circle of (e) and (h).}
       \label{Halo}
\end{figure}
The setup (Fig.\ref{Setup}) uses for $LP_{11}$ injection the lateral refraction in an embedded fiber as described in \cite{kalaidji09}. The oblong coiling is preceded by an arbitrary length of fiber that can be slightly curved and deliberately followed by a long coil ($\approx 40 m$). To compensate these perturbations, we introduced a quarter-wave and half-wave plate at the laser output, and a polarization controller ({\it Newport F-POL-IL}). Great care was taken to avoid twisting the fiber and to maintain it in a horizontal plane as good as possible. The polarization of the output beam is measured by a rotating analyzer. Fig.\ref{Halo} shows the output pattern for 8 successive angles of the analyzer with a $\pi /8$ increment and some symmetry criteria that are extracted from a simple image analysis: the arrows point the analyzer position, whereas dotted lines are the symmetry axis passing through the maximums of the two lobes, and the best line that is deduced by a mean square calculation from the loci of the minimums between the two lobes. They are orthogonal with an accuracy of 2.5 degrees. Dash-and-dot lines are deduced from the axis of the reference pattern (b) by the same rotation $\alpha$ as the analyzer.
Our adjustment criteria were only minimizing the power modulation and the existence of a perfectly dark line between the two lobes during the whole analyzer rotation. This result exhibits an in-phase combination of $TE_{01}$ and $TM_{01}$ that can be easily converted to $TM_{01}$ by two half-wave plates \cite{grosjean05}. It was obtained by an empirical adjustment of the polarization controller and of the translation stage. The difference of a few degrees between the analyzer rotation and the lobe rotation as well as a small relative power modulation of $\pm 13 \%$ (i) indicate a residual presence of $LP_{11}$ mode. These defaults are amplified by the presence of a small amount of fundamental mode, because the refraction component was built on a too short fiber segment that was spliced to a longer one. This leads to a slight dissymmetry in the pattern without analyzer (i), an imbalance (k) or a bridge between the two lobes for some azimuthal position (a, d, h). The difficulty of obtaining a stable pattern in the long term can be attributed to the long coil, to the imperfect attachments of the fiber and its friction on the cylinders.

The oblong coiling demonstrates the possibility of selecting a single doughnut mode with an all-fiber device. Some limitation are due to the fiber length that was chosen: a complementary couple of crystalline plates is needed at the output. The efficiency of a twisted segment could
be tested to compensate the residual twist. The adjustment facility would be improved by a real time pattern analysis and a better mechanical control of fiber tension. Special fibers with two concentric cores \cite{ramachandran05} demonstrated their ability to lift polarization degeneracy, hence to reduce the beat length of doughnut modes to less than one centimeter. They will improve the stability of the selected mode along the guided propagation and the polarization preservation. Moreover it opens the possibility of using a series of a few ridges if the leakage induced by the high curvature of the oblong coiling becomes prohibitive.

\end{document}